\documentclass[12pt]{article}

\usepackage{epsfig}
\usepackage{graphicx}
\usepackage{amsmath}
\usepackage{amssymb, verbatim}
\usepackage{fullpage}

\begin{document}

%%%%%%%%% TITLE
\title{Automatically Creating Design Models from 3D Anthropometry Data}

\author{Stefanie Wuhrer\footnote{National Research Council of Canada} \footnote{Saarland University, Germany} \footnote{Max-Plank Institute Informatik, Germany} \and Chang Shu$^*$ \and Prosenjit Bose\footnote{Carleton University, Canada}}

\maketitle

\begin{abstract}
When designing a product that needs to fit the human shape, designers often use a small set of 3D models, called {\em design models}, either in physical or digital form, as representative shapes to cover the shape variabilities of the population for which the products are designed. Until recently, the process of creating these models has been an art involving manual interaction and empirical guesswork. The availability of the 3D anthropometric databases provides an opportunity to create design models optimally. In this paper, we
propose a novel way to use 3D anthropometric databases to generate design models that represent a given population for design applications such as the sizing of garments and gear. We generate the representative shapes by solving a covering problem in a parameter space. Well-known techniques in computational geometry are used to solve this problem. We demonstrate the method using examples in designing glasses and helmets.
\end{abstract}

%--------------------------------------------------------------------------------------------------------------------------------------------------------------------------------------------------------
\section{Introduction}

In designing wearable products, such as garments or head gears, one of the most important objectives is to make the products to fit the humans comfortably. In order to accommodate the human shape variabilities, different sizes are usually created. Traditionally, this is achieved by considering a few key dimensions of the human body. To design a sizing system, anthropometric measurement data of these dimensions are collected 
and tabulated. Then, a griding of the dimensions is formed to create the sizes~\cite{kwon-et-al-09}~\cite{Roebuck-Kroemer-Thomson-75}.

Although traditional anthropometry has a long history and has accumulated a vast amount of data, it is limited by its tools (mainly tape measure and caliper); the sparse dimensional measurements do not provide sufficient shape information. In many applications, designers need full 3D models as representative shapes for the sizes~\cite{Roebuck-Kroemer-Thomson-75}.
These models are sometimes called manikins when the full body is concerned, or head forms in case of the head and face are concerned. For the purpose of generality, we call them {\em design models} in this paper. These 3D models provide the overall shape of the human body. They need to be chosen carefully to ensure that the manufactured items fit the target population.

Without 3D information, design models are often created by artists who sculpt out the 3D forms, interpolating the measurement dimensions using their experience in creating human shapes~\cite{meunier_shu_xi_09}. These approaches are labour-intensive and do not create accurate human models.

3D anthropometric data, obtained using 3D imaging technologies, provide detailed shape information. In addition, traditional measurements can also be extracted from the 3D models. Therefore, 3D anthropometric data offers an opportunity to improve the quality of the design models and, at the same time, maintain the simplicity of the traditional design schemes where key body dimensions are used.

Despite the fact that 3D anthropometric databases have been available for two decades, surprisingly little research has been reported concerning using the 3D data to improve the design models. Robinette~\cite{robinette_3D_design}, Ball~\cite[Chapter 7]{roger_ball_thesis}, and Meunier et al.~\cite{meunier_shu_xi_09} are recent attempts in this direction. However, all of these methods rely heavily on manual interaction and do not give a systematic way to optimize the design models such that the sizing system provide maximal accommodation. 

In this paper, we present a method for automatically creating a set of design models based on a 3D anthropometric database and optimally computing the design models that represent the underlying target population. 
Starting with a 3D anthropometric database $X$ that represents the target population for the product to be designed,  the method requires as input the ordered set of $d$ measurements $M$ on the human body that need to be considered during the design phase. For instance, when designing glasses, the set $M$ to be considered are the width of the face and the width of the bridge of the nose. Furthermore, the method requires as input an ordered set of $d$ tolerances $T$; one tolerance for each measurement in $M$.
In the example with the glasses, we may know that the glasses to be designed can be adjusted by $2.76cm$ to fit faces with different widths and by $0.19cm$ to fit noses with different widths of the bridge. 
In this case, $T$ contains the tolerances $2.76$ and $0.19$. 
If the product is non-adjustable, the tolerances represent the measurement intervals that the sizes cover.
With this information, the method finds a set of design models that optimally represent the population acquired in the database for the specific purpose of designing a product that depends on the measurements $M$ and that allows adjustments along the tolerances $T$.

The method proceeds by mapping each of the subjects in the database to a point in $\mathbb{R}^d$ and by solving a discrete covering problem in this space. We use the subjects in the database directly instead of learning their underlying probability distribution since the type of distribution is usually unknown. The design models created using this approach are expected to have higher data accuracy and completeness than traditionally produced design models because we compute the best design models from a database that represents the target population.

Sections~\ref{conversion},~\ref{cover_box}, and~\ref{fixed_box} discuss how to convert the problem to a discrete covering problem and how to solve the covering problem. To obtain the design models, the method employs the full Procrustes mean~\cite{dryden_mardia_shape_analysis} as outlined in Section~\ref{gpa}. The design models represent the given database. If the database is large, we expect good design models. If the database contains few subjects, it is possible to assume and fit a probability distribution satisfied by the population and to extrapolate subjects from the database based on that distribution. In case of a small database, this approach may provide more accurate results. Section~\ref{extrapolate} discusses this option. Section~\ref{experiments} gives design examples.

Some prior work has focused on operating in a parameter space based on traditional sparse anthropometric measurements. If only sparse anthropometric measurements are known, it is not straight forward to find 3D design models because computing a 3D model based on a small set of measurements is an under constrained problem. In this work, the shape information is provided by a 3D anthropometric database. This data allows to compute design models by constraining them to lie within the shape space spanned by the database.

In this work, we use the Civilian American and European Surface Anthropometry Resource (CAESAR) database~\cite{robinette_daanen_paquet_99_caesar} to compute a sizing system. This database contains human shapes in a similar posture. Furthermore, each shape contains a set of 73 anthropometric landmarks. We exploit this information to parameterize the models.

%--------------------------------------------------------------------------------------------------------------------------------------------------------------------------------------------------------
\section{Related Work}

McCulloch et al.~\cite{mcculloch_paal_ashdown_98} used the traditional sparse anthropometric measurements to automatically create an apparel sizing system that has good fit. For a fixed set of measurements on the body, a fixed number of sizes, and a fixed percentage of the population that needs to be fitted by the sizing system, they optimize the fit of the sizing system. The fit for a human body is defined by a weighted distance function between the measurements on the body and the measurements used for the sizing system. Optimizing the fit amounts to solving a non-linear optimization system. This system is hard to solve and suboptimal solutions may be obtained. This approach can operate in multi-dimensional spaces. It is not straight forward to obtain design models from the resulting sizing system because computing a design model from a sparse set of measurements is an under constrained problem.

Mochimaru and Kouchi~\cite{mochimaru_kouchi_00} represented each model in a database of human shapes using a set of manually placed landmark positions. They proposed an approach to find representative three-dimensional body shapes. The approach first reduces the dimensionality of the data using multi-dimensional scaling, and then uses Principal Component Analysis (PCA) to find representative shapes. Mochimaru and Kouchi showed that this approach is suitable to find representative shapes of a human foot. While this approach is fully automatic, it assumes that the distortion introduced by multi-dimensional scaling is small. While this may be true for low-dimensional data, it is not in general true when three-dimensional measurements of high resolution are considered. Hence, there is no guarantee that the design models optimally represent the population.

Recently, three-dimensional measurements of high resolution have been used to aid in the design process. Robinette~\cite{robinette_3D_design} studied different ways to align a database of 3D laser scans of heads for helmet design. The goal is to find alignments with minimum variability of shape. This way, one can design helmets that offer maximum protection. Meunier et al.~\cite{Meunier2000361} used a database of 3D laser scans of heads to objectively assess the fit of a given helmet. This assessment strategy is useful to analyze existing designs, but it does not produce design models. Kouchi and Mochimaru~\cite{kouchi_mochimaru_04} proposed an approach to design spectacle frames based on database of human face shapes. The approach proceeds by analyzing the morphological similarities of the faces in the database and by dividing the faces into four groups. For each group, a representative form is found automatically. In this work, a spectacle frame was designed for each representative shape, and it was shown that a good fit was achieved. Guo et al.~\cite{guo_et_al_09} proposed a similar approach to construct design models for helmet design using a database of magnetic resonance images of heads. The approach proceeds by automatically dividing the head shapes into groups and by constructing one design model per group. Although the methods by Kouchi and Mochimaru and Guo et al. require little manual work, the methods are specific to spectacle and helmet design, respectively, and cannot easily be extended to the design of other gear or garments. Furthermore, the division of heads into groups is not guaranteed to produce design models that optimally represent a target population.

Meunier et al.~\cite{meunier_shu_xi_09} propose to parameterize a set of head scans and to perform PCA on the parameterized scans. They then use the first two principal components to manually design a set of three design models. While this approach aims to exploit the information provided by a parameterized database of head shapes, the approach is heuristic and assumes that the data follows a Gaussian distribution. Furthermore, manually picking the design models based on a learned distribution is hard in a high-dimensional space without reducing the dimensionality of the data because humans cannot easily visualize high-dimensional spaces. Hence, if the aim is to consider four or more principal components, the approach by Meunier et al. becomes very difficult to use.

We propose a fully-automatic method to compute design models that represent a given 3D anthropometric database well. Since the approach computes the design models automatically, it can operate in high-dimensional spaces. Unlike the method by McCulloch et al.~\cite{mcculloch_paal_ashdown_98}, our method does not rely on solving a non-linear optimization system. Instead, we model the fit explicitly using a set of tolerances (one tolerance along each dimension) that explains by how much the garment or gear can be adjusted along each dimension. These tolerances depend directly on the design and the materials that are used in a specific application. In this way, our method finds optimal design models for a specific task such as helmet design. To our knowledge, this is the first method that simultaneously considers the optimal fit accommodation and design models.

%--------------------------------------------------------------------------------------------------------------------------------------------------------------------------------------------------------
\section{Converting 3D Shape Data Into Parameter Space Points}
\label{conversion}

The proposed approach proceeds by first parameterizing all of the $n$ subjects present in the database $X$. In general, finding accurate point-to-point correspondences between a set of shapes is a hard problem~\cite{vanKaick_egstar10}. However, in our application, we know that the shapes are human body shapes, which allows the use of template-based approaches to parameterize the database. 

First, consider the case where the scans are assumed to be in similar posture. This assumption is common in our application because in a typical 3D anthropometry survey, the human subjects are asked to maintain a standard posture. In this case, we can proceed by first using Ben Azouz et al.'s~\cite{benazouz_shu_mantel_06_landmarks} approach to automatically predict landmarks on the scans followed by Xi et al.'s approach~\cite{xi_lee_shu_07_bodies} to parameterize the models. Ben Azouz et al.'s approach proceeds by learning the locations of a set of anthropometric landmarks from a database of human models using a Markov Random Field (MRF), and by using probabilistic inference on the learned MRF to automatically predict the landmarks on a newly available scan. Xi et al.'s approach~\cite{xi_lee_shu_07_bodies} exploits the anthropometric landmarks to fit a template model to the scans. The method proceeds in two steps. First, it computes a radial basis function that maps the anthropometric landmarks on the template mesh to the corresponding landmarks on the scan. This function is used to deform the template mesh. Second, the method further deforms the template to fit the scan using a fine fitting approach as in Allen et al.~\cite{allen_curless_popovic_03_parametrization_body_shape}. 

Second, consider the general case where the postures of the scans vary. In this case, we can again proceed by first predicting landmarks on the scans automatically and by using a template-based deformation to parameterize the models. When using this method~\cite{wuhrer_shu_xi_vc10}, the locations of the landmarks are learned and predicted in an isometry-invariant space and the template fitting approach uses a skeleton-based deformation to allow for posture variation.

In this work, we use the CAESAR database. The models of this database are in similar posture and for each model, we know a set of 73 manually placed anthropometric landmarks. Hence, we use the anthropometric landmarks as input to Xi et al.'s approach~\cite{xi_lee_shu_07_bodies} to parameterize the models.

Once the database is parameterized, each subject $X_i$ of the database is represented by a triangular mesh. We can now measure corresponding distances on all of the bodies. This paper considers Euclidean distances. However, the techniques outlined in the following extend to arbitrary distances, such as geodesics. 

The approach computes the ordered set $M$ of $d$ distances that are meaningful for the design of a specific garment or gear for each of the $n$ bodies in the database. For each body, the $d$ measurements in order can be viewed as a point $P_i$ in the \textit{parameter space} $\mathcal{S} = \mathbb{R}^d$. The entire database is then represented as a set $P$ of $n$ points $P_0, P_1,\ldots, P_{n-1}$ in $\mathcal{S}$. 

The designer can specify a range of fit along each of the $d$ dimensions that were measured on the bodies. This range specifies how much a garment stretches or by how much gear can be adjusted in the given direction. The ordered set $T$ of $d$ ranges in order defines the side lengths of a $d$-dimensional box in $\mathcal{S}$. Let $s_i(B)$ denote the side length in dimension $i$ and let $B$ denote the $d$-dimensional box with side lengths $s_i(B)$ centered at the origin.

The problem of computing a sizing system that fits the given population can now be expressed as a covering problem in $\mathcal{S}$. That is, we aim to cover all of the points in $P$ with translated copies of $B$. Sections~\ref{cover_box} and~\ref{fixed_box} discuss how to solve this problem.

Once the points in $P$ are covered by a set of translated copies of $B$, we aim to convert each box $B_i$ back into a body shape that represents the points covered by $B_i$. This is achieved by computing the full Procrustes mean shape~\cite{dryden_mardia_shape_analysis} of the body shapes corresponding to the points covered by $B_i$. Section~\ref{gpa} discusses this step in detail.

%--------------------------------------------------------------------------------------------------------------------------------------------------------------------------------------------------------
\section{Covering Parameter Space With Boxes}
\label{cover_box}

This section discusses the problem of finding the minimum number of translated copies of $B$ that cover all of the points in $P$. 

We first analyze how many translated copies of $B$ need to be considered by the algorithm. When considering only one distance dimension, the box $B$ becomes a line segment of fixed length. Note that two line segments that are combinatorially equivalent (i.e. they cover the same points of $P$) do not add anything to the space of solutions. Therefore, the algorithm only needs to consider translated line segments that are combinatorially different from each other. Since the set covered by a translated copy of $B$ only changes if a point of $P_i$ either enters or leaves the set, the algorithm only needs to consider the translated line segments with endpoints $P_i$. Since a line segment has two endpoints and since there are $n$ points in $P$, we need to consider $2n$ boxes. Extending this argument to $d$ distance dimensions yields that at most $(2n)^d$ translated copies of $B$ need to be considered.

The problem of finding the minimum number of boxes among the set of at most $(2n)^d$ translated copies of $B$ that cover all of the points in $P$ can now be viewed as a set cover problem. A history of this problem can be found in Vazirani~\cite[Chapter 2]{vazirani_01}. This problem is NP-hard~\cite{karp} and it is therefore impractical to find an optimal solution. Hence, we aim to find an approximation to the solution that has bounded error. A solution is a $(1+\epsilon)$-approximation of the optimal solution if it uses $k'$ boxes with $k' \leq (1+\epsilon)k$ for any $\epsilon > 0$, where $k$ is the number of boxes used by the optimal solution. In the following, we discuss how to compute a $(1+\epsilon)$-approximation~\cite{hochbaum_maass_covering}. We then discuss a more efficient greedy algorithm to find a set of boxes that is a $(2d\log(n))$-approximation of the optimal solution.

\subsection{$(1+\epsilon)$-Approximation}

This section summarizes the approach by Hochbaum and Maass~\cite{hochbaum_maass_covering} to find a $(1+\epsilon)$-approximation to the covering problem. The approach proceeds by dividing the parameter space into a regular grid with width $s_i(B)$ along each dimension. For a given input parameter $l$ related to $\epsilon$ $((1+\epsilon) = (1+\frac{1}{l})^d)$, the approach partitions the parameter space into slabs of $l$ grid cells along each dimension. Note that $l^d$ partitions are possible because the locations of the slabs can be shifted $l$ times along each dimension. For each of the partitions, the algorithm computes the union of optimal solutions in the sets of $\underbrace{l\times l \times \ldots \times l}_{d\mbox{ times}}$ grid cells created by the partition. The algorithm takes $O(l^d n^{2l^d+1})$ time. For a detailed analysis, refer to~\cite{hochbaum_maass_covering}. This is only practical for problems with few points in low dimensions. Hence, we only implemented this approach for $d=2$.

\subsection{Greedy Covering}
\label{greedy_cover}

This section summarizes an efficient algorithm to find a solution to the covering problem. Unfortunately, the solution is not guaranteed to be a $(1+\epsilon)$-approximation of the optimal solution. The approach proceeds by considering $r$ translated copies of $B$. We denote these boxes $B_0,B_1,\ldots,B_{r-1}$. For each box $B_i$, we compute the set of points of $P$ covered by $B_i$. This yields a collection of $r$ sets. Our goal is to find a small subset of the sets that covers all of the points in $P$. We solve the problem using a greedy approach by repeatedly selecting the set covering the maximum number of uncovered points until all points are covered. Finding the set of points covered by one box $B_i$ takes $O(dn)$ time. Hence, finding all of the sets takes $O(dnr)$ time. We store the following information. For each point $P_i$, we store which boxes contain $P_i$, for each box $B_i$, we store which points are contained in $B_i$, and we store for each box $B_i$ the number of uncovered points in $B_i$. The set that covers the maximum number of uncovered points can now be found in $O(r)$ time. Furthermore, using the stored information, we can remove one point from all of the boxes $B_i$ in $O(r)$ time. Since there are a total of $n$ points, the algorithm takes $O(nr)$ time. Hence, this approach takes $O(dnr)$ time.

As discussed above, if we consider all of the combinatorially different boxes $B_i$, $r$ is at most $(2n)^d$. When all combinatorially different boxes are considered, it can be proven that this greedy approach is a $\log(n)$-approximation of the optimal solution~\cite[Chapter 2]{vazirani_01}. Hence, we can find a $\log(n)$-approximation of the optimal solution in $O(dn^{d+1})$ time. 

Denote the set of all combinatorially different boxes by $\mathcal{B}^*$. In practice, we reduce the number of boxes further by considering only the translated copies of $B$ with centers in $P$. Denote the set of these boxes by $\mathcal{B}$. This way, we obtain $r=n$ and reduce the running time to $O(dn^2)$ at the cost of not considering all of the combinatorially different boxes. To analyze the approximation ratio of this solution, consider the case $d=1$, where the boxes become line segments. Assume that the optimal solution picks a line segment $B_i^*$ not included in $\mathcal{B}$. Recall that $B_i^*$ is in the set $\mathcal{B}^*$. All of the elements covered by $B_i^*$ can be covered by two line segments in $\mathcal{B}$; namely, the line segment centered at the leftmost point in $B_i^*$ and the line segment centered at the rightmost point in $B_i^*$. This argument can be generalized to $d$-dimensional space as follows. A box $B_i^*$ picked by the optimal solution that is not included in $\mathcal{B}$ can be covered by $2d$ boxes in $\mathcal{B}$. Hence, the optimal solution using $\mathcal{B}$ needs at most $2d$ times the number of boxes picked by the optimal solution using $\mathcal{B}^*$. Since we solve the problem using the greedy algorithm on $\mathcal{B}$ and since the greedy algorithm is known to compute a $\log(n)$-approximation of the optimal solution~\cite[Chapter 2]{vazirani_01}, our approach is guaranteed to compute a $(2d\log(n))$-approximation of the optimal solution.

%--------------------------------------------------------------------------------------------------------------------------------------------------------------------------------------------------------
\section{Covering With Fixed Number of Boxes}
\label{fixed_box}

In some applications such as the design of garments, it is not desirable to produce sizes for an entire population. Instead, one wishes to produce a fixed number $k$ of sizes that fits the largest portion of the population. For instance, a company that manufactures T-Shirts may wish to manufacture three sizes (small, medium, and large) in a way that the sizes fit the largest possible portion of the population. 

In parameter space, this means that we do not wish to cover all of the points in $P$. Instead, the goal is to cover the maximum number of points in $P$ with a fixed number $k$ of boxes. This problem is also NP-hard since a polynomial-time algorithm to solve this problem would give a polynomial-time algorithm to solve problem considered in Section~\ref{cover_box}. Hence, we solve this problem using a greedy approach. Unfortunately, the solution is not guaranteed to be a $(1+\epsilon)$-approximation of the optimal solution. However, by a similar argument to the one in the previous section, we can show that the approach computes a $(2d\log(n))$-approximation of the optimal solution. The greedy approach proceeds as in Section~\ref{greedy_cover}. The only difference is that we stop after the first $k$ boxes are selected. Using an analysis similar to the one in Section~\ref{greedy_cover}, it can be shown that this algorithm takes $O(dnr)$ time, where $r$ is the number of boxes considered by the algorithm. Recall that $r=O((2n)^d)$ in theory and that we set $r=n$ in practice.

%--------------------------------------------------------------------------------------------------------------------------------------------------------------------------------------------------------
\section{Computing Representative Shapes for Boxes}
\label{gpa}

Once the algorithm selected a set of $k$ boxes $B_0,B_1,\ldots,B_{k-1}$ to represent the given measurements, we aim to convert each box $B_i$ back into a body shape that represents the points covered by $B_i$. This is achieved by finding the parameterized body shapes corresponding to points in $P$ covered by $B_i$ and by computing the full Procrustes mean of these shapes~\cite{dryden_mardia_shape_analysis}. To compute the full Procrustes mean of $m$ shapes, we repeatedly compute the average of the $m$ shapes and align each of the $m$ shapes to the average shape using a rigid transformation.

If the box $B_i$ contains a sufficient number of points and if the mean of these points is close to the center of $B_i$, the Procrustes mean is a good representative of $B_i$. Otherwise, the Procrustes mean of these shapes will not yield a good representation of the shapes covered by $B_i$. If this situation occurs, the approach can be modified by sampling a set of points $P_{new}$ in $B_i$ and by finding the shapes $X_{new}$ corresponding to these points as outlined in Section~\ref{extrapolate}. We can then find a design model by computing the full Procrustes mean of the shapes $X_{new}$.

This approach yields a set of $k$ body shapes $S_i$ corresponding to the boxes $B_i$. The shapes $S_i$ can now be used for the design of garments or gear. Note that the shapes $S_i$ are not simply scaled versions of each other since each shape $S_i$ is derived from a different set of true body scans. We use each shape $S_i$ as a design model.

%--------------------------------------------------------------------------------------------------------------------------------------------------------------------------------------------------------
\section{Extrapolating Shapes from the Database}
\label{extrapolate}

The design models computed in the previous sections represent the given database. If the database contains few subjects, the coverage of the target population by the computed design models may be small. In this case, extrapolating subjects from the database may provide more accurate results. In order to extrapolate subjects from the database, we need to assume that the target population obeys a specific probability density. In this section, we assume that the data follows a Gaussian distribution. 

We discuss how to extrapolate subjects from the database to improve the coverage of our method. We use the approach by Allen et al.~\cite{allen_curless_popovic_03_parametrization_body_shape} to compute a new shape $X_{new}$ based on a new point $P_{new}$ in $\mathcal{S}$. Since the database is parameterized and follows a Gaussian distribution, we can perform PCA of the data. In PCA space, each shape $X_i$ is represented by a vector $W_i$ of PCA weights. PCA yields a mean shape $\mu$ and a matrix $A$ that can be used to compute a new shape $X_{new}$ based on a new vector of PCA weights $W_{new}$ as $X_{new} = A W_{new} + \mu$. Recall that we aim to create a new shape based on a new point $P_{new}$ in $\mathcal{S}$. To achieve this goal, we use the database to learn a linear mapping from $P_i$ to $W_i, i=0,\ldots,n-1$. This mapping is called \textit{feature analysis} and described in detail by Allen et al.~\cite{allen_curless_popovic_03_parametrization_body_shape}. Feature analysis yields a matrix $B$ that can be used to compute a new vector of PCA weights $W_{new}$ based on a new point $P_{new}$ as $W_{new}=B P_{new}$. This model allows to compute a shape $X_{new}$ based on a new point $P_{new}$ as $X_{new} = A B P_{new} + \mu$. 

It remains to outline how to find new points $P_{new}$. We assume that the points $P_i, i=0\ldots,n-1$ can be modeled by a Gaussian distribution. We learn the distribution from the given database using maximum likelihood estimation. We then sample a set of points from this distribution and find the corresponding shapes using feature analysis. We use these new subjects along with the given database to find the design models. 

Another way to find new points $P_{new}$ is to sample the surface of a fixed equal probability density of the learned Gaussian distribution (this surface is an ellipsoid). This approach is useful to increase boundary coverage.

\begin{figure}[htb]
\centering
\includegraphics[width = 5.0cm]{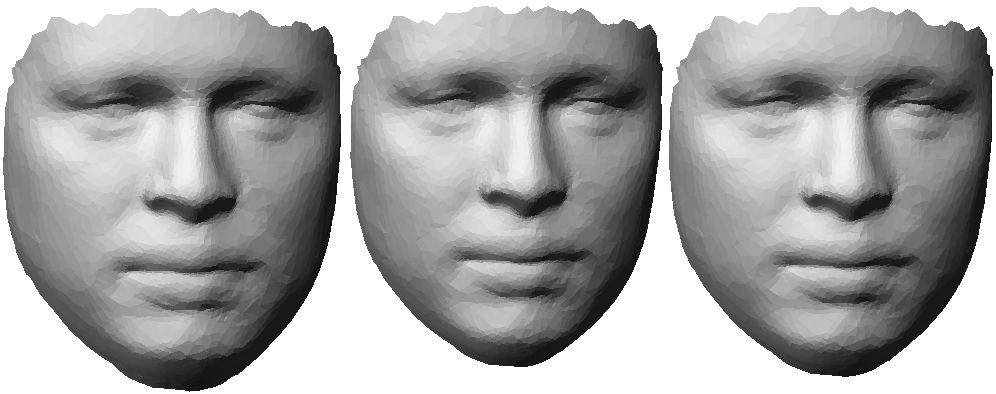}
\caption{\textit{Faces obtained using feature analysis.}}
\label{extra_faces}
\end{figure}

We use as database 50 faces containing 6957 triangles each from the CAESAR database~\cite{robinette_daanen_paquet_99_caesar}. Figure~\ref{extra_faces} shows three faces that were obtained using feature analysis.

%--------------------------------------------------------------------------------------------------------------------------------------------------------------------------------------------------------
\section{Examples}
\label{experiments}

We demonstrate the proposed approach for the design of glasses and for the design of helmets. The first example shows the concept of the presented approaches using a small database of faces. The second example gives an evaluation of the greedy approaches using a large database of heads.

\subsection{Design of glasses}

\begin{figure}[htb]
\centering
\includegraphics[width = 2.5cm]{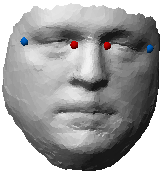}
\caption{\textit{Measurements used to define parameter space.}}
\label{measurements}
\end{figure}

When designing glasses, one may wish to measure the width of the face and the width of the bridge of the nose. Using these two measurements yields $\mathcal{S} = \mathbb{R}^2$. Figure~\ref{measurements} shows the measurements on one face in the database. The first dimension measures the Euclidean distance between the blue points and the second dimension measures the Euclidean distance between the red points.

For this example, we use 50 faces from the CAESAR database~\cite{robinette_daanen_paquet_99_caesar}. We choose a small database since this allows to illustrate the result of the $(1+\epsilon)$-approximation.

In our example, we aim to design glasses that can be adjusted by $2.67cm$ in the first dimension and by $0.19cm$ in the second dimension. This defines the box $B$.

\begin{figure*}[htb]
\centering
\includegraphics[width = 10.0cm]{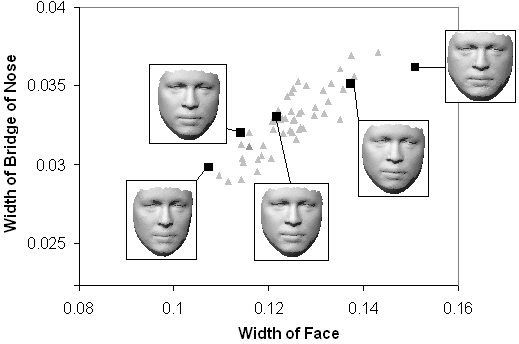}
\caption{\textit{Covering parameter space with $\epsilon = 1.25$. Unit along axes is meters.}}
\label{approx}
\end{figure*}

Figure~\ref{approx} shows the result when covering $P$ using a $(1+\epsilon)$-approximation with $\epsilon = 1.25$. The figure shows the points $P$ as grey triangles and the centers of the boxes $B_i$ as black squares. Furthermore, for each box, the figure shows a screen shot of the corresponding Procrustes mean shape. These face shapes can be used to create the sizes for the glasses.

\begin{figure*}[htb]
\centering
\begin{tabular}{c c}
\includegraphics[height = 6.7cm]{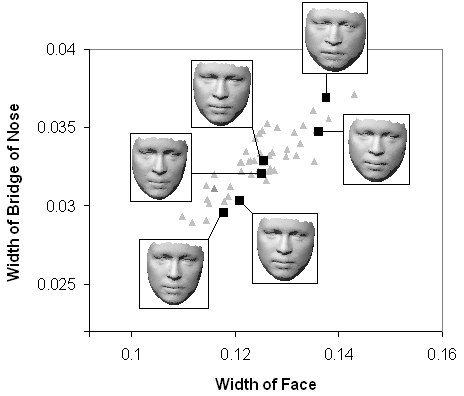} \hspace{0.5cm}&
\hspace{0.5cm} \includegraphics[height = 6.0cm]{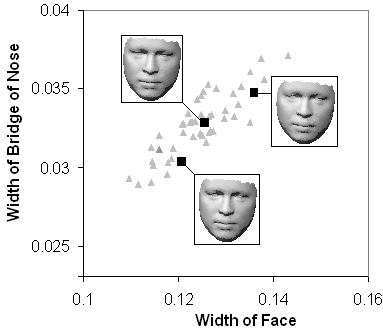} \\
(a) \hspace{0.5cm} & \hspace{0.5cm} (b) \\
\end{tabular}
\caption{\textit{Greedy covering of parameter space. (a): Complete covering. (b): Covering with three boxes. Unit along axes is meters.}}
\label{greedy}
\end{figure*}

Figure~\ref{greedy} shows the result of a greedy covering of $P$. Recall that this is a $(4\log(n))$-approximation of the optimal solution. The symbols used in the figure are identical to the ones in Figure~\ref{approx}. Figure~\ref{greedy}(a) shows the result when we aim to cover the entire population. We can see that in this example, we only require one extra shape when covering with the greedy algorithm than when covering using a $(1+\epsilon)$-approximation with $\epsilon = 1.25$. Figure~\ref{greedy}(b) shows the result of greedily covering a large subset of $P$ with three boxes. We can see that the selected boxes cover the parameter space well.

\subsection{Design of helmet}

\begin{figure}[htb]
\centering
\includegraphics[width = 5.5cm]{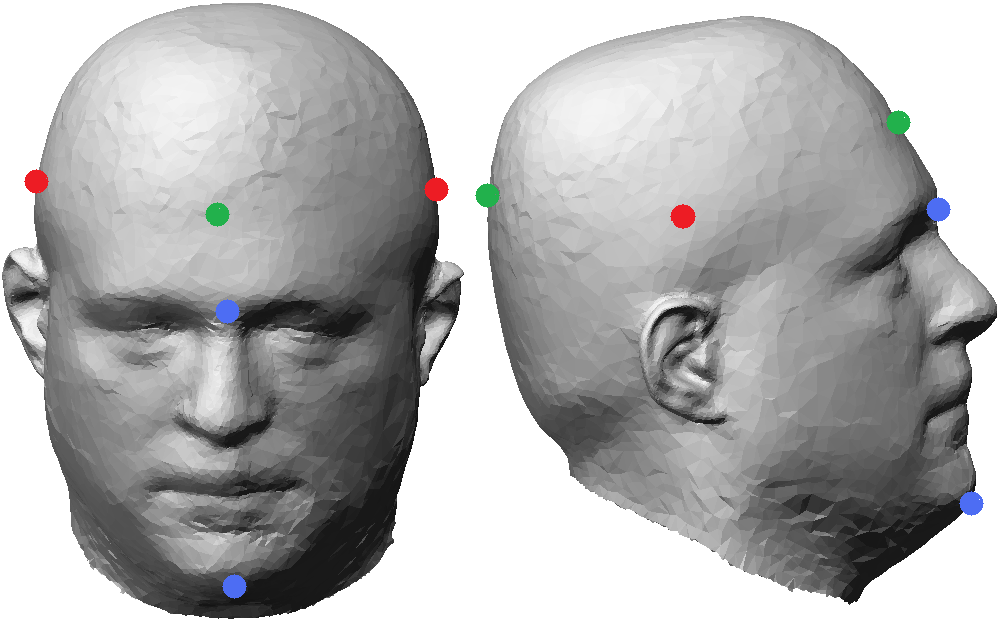}
\caption{\textit{Measurements used to define parameter space.}}
\label{measurements_helmet}
\end{figure}

When designing a helmet, the three most crucial measurements are the head width, the head depth, and the face hight~\cite{meunier_shu_xi_09}. Using these three measurements yields $\mathcal{S} = \mathbb{R}^3$. Figure~\ref{measurements_helmet} shows the measurements on one head in the database. The first dimension measures the Euclidean distance between the red points, the second dimension measures the Euclidean distance between the green points, and the third dimension measures the Euclidean distance between the blue points.

For this example, we conduct an evaluation. We use 1500 heads from the CAESAR database~\cite{robinette_daanen_paquet_99_caesar} to compute the design models and we then test the quality of fit using 500 different heads from the CAESAR database. 

In our example, we aim to design helmets that can be adjusted by $2.99cm$ in the first dimension, by $3.25cm$ in the second dimension, and by $3.33cm$ in the third dimension. This defines the box $B$.

\begin{figure*}[htb]
\centering
\includegraphics[width = \textwidth]{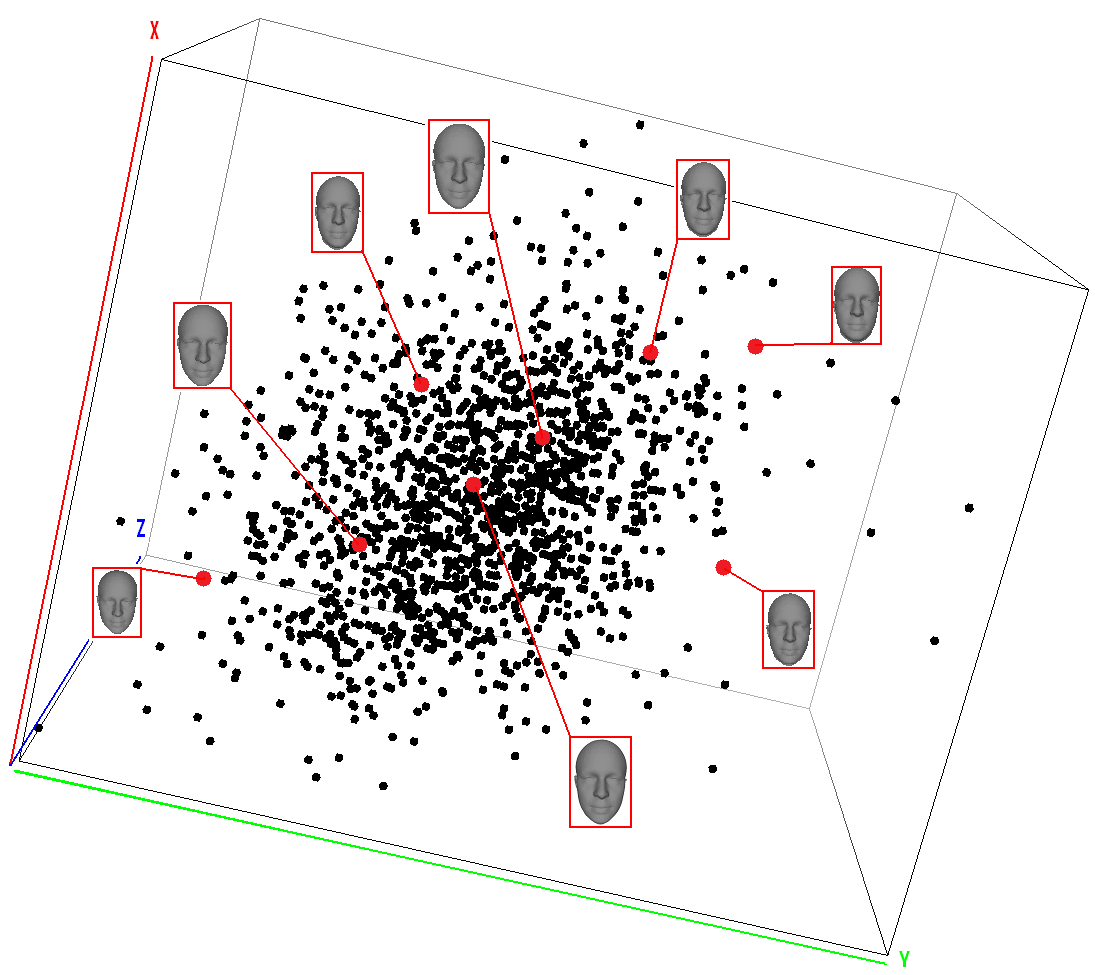} 
\caption{\textit{Greedy complete covering of parameter space.}}
\label{greedy_head_complete}
\end{figure*}

\begin{figure*}[htb]
\centering
\includegraphics[width = 16.0cm]{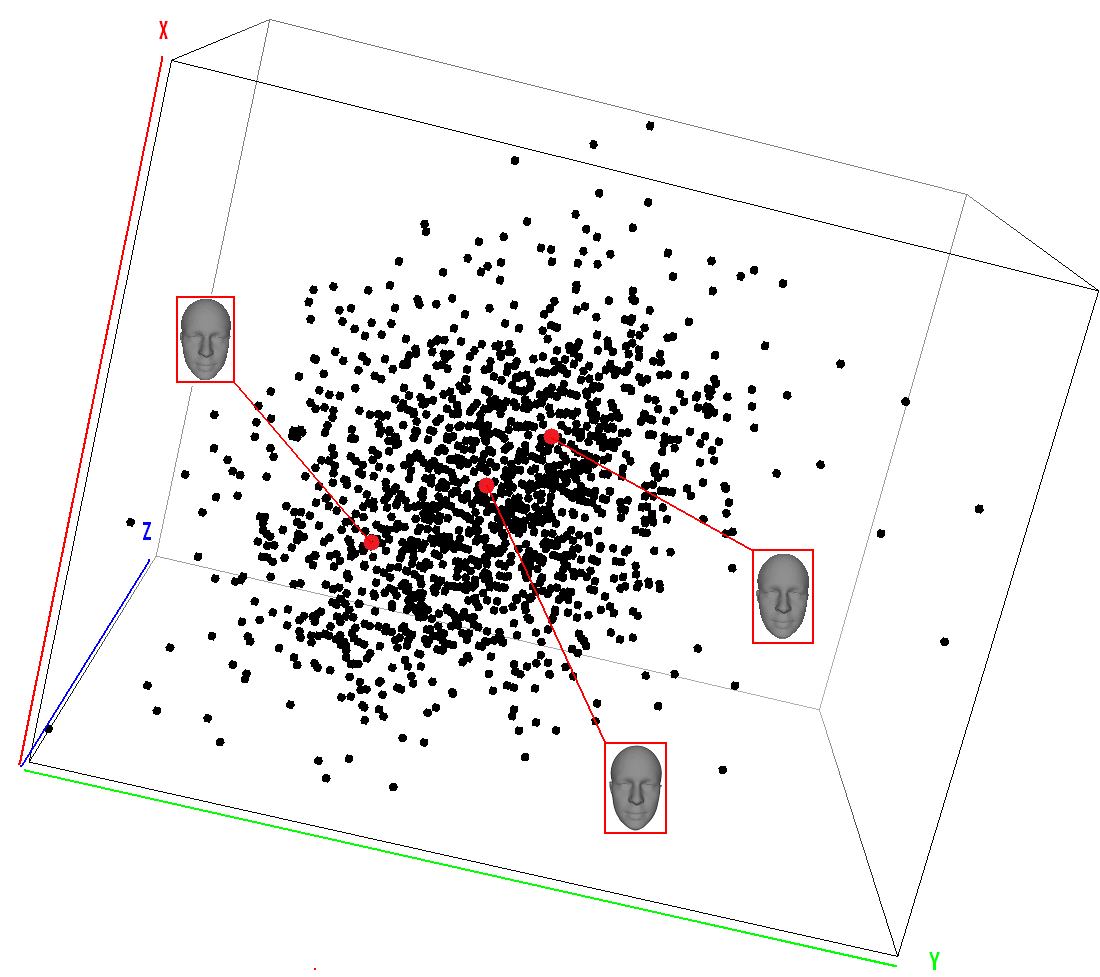} \\
\caption{\textit{Greedy covering of parameter space with three boxes.}}
\label{greedy_head}
\end{figure*}

Figures~\ref{greedy_head_complete} and~\ref{greedy_head} show the results of a greedy covering of $P$. The figures show the points $P$ as black points and the centers of the boxes $B_i$ as red points. Furthermore, for each box, the figures show a screen shot of the corresponding Procrustes mean shape. The coordinate axes are shown in the colour of the corresponding dimension (see Figure~\ref{measurements_helmet}). Figure~\ref{greedy_head_complete} shows the result when we aim to cover the entire population. This covering requires eight design models. Figure~\ref{greedy_head} shows the result of greedily covering a large subset of $P$ with three boxes. The boxes corresponding to the three design models cover a large subset of $P$. Note that for the data used in this example, it is not easy to manually find the best locations of the boxes since we use three measurements and since it is hard to optimally place points manually in three-dimensional space.

We use 500 different head shapes to compute the quality of fit of the computed design models as follows. We compute the points in $\mathcal{S}$ corresponding to the 500 head shapes and we compute how many of these points are covered by at least one of the computed boxes. The eight design models computed using the greedy covering algorithm shown in Figure~\ref{greedy_head_complete} cover $99.8\%$ of all the shapes. The three design models computed by greedily covering the largeset subset of $P$ using three boxes shown in Figure~\ref{greedy_head} cover $95.2\%$ of all the shapes. This shows that when using the three design models shown in Figure~\ref{greedy_head} to design three sizes of a helmet, we expect that $95.2\%$ of all adults find that at least one of these three helmets fits them.

%--------------------------------------------------------------------------------------------------------------------------------------------------------------------------------------------------------
\section{Conclusion}

This paper presented a novel approach to generate design models for the design of gear or garments. The approach makes use of the widely available anthropometric databases to find design models that represent a large portion of the population. We find the design models by solving a covering problem in a low-dimensional parameter space. 

Note that in this paper, we use translated boxes of the same size to cover the parameter space. This seems the most intuitive shape with which a designer may wish to cover the parameter space. To solve this problem, we presented a slow $(1+\epsilon)$-approximation and a fast $(2d\log(n))$-approximation of the optimal solution. If the aim is to cover the parameter space with translated balls or ellipsoids of the same size, all of the algorithms in this paper can be adapted to this scenario. If the aim is to cover the parameter space with boxes, balls, or ellipsoids of different size or orientation, the algorithms presented in Sections~\ref{greedy_cover} and~\ref{fixed_box} can be adapted to this scenario.

The resulting design models are only as good as the given anthropometric database. This paper discussed the option of improving the coverage of the target population by extrapolating models from the database before computing the design models.

\section*{Acknowledgment}

We thank Pengcheng Xi for providing us with the data. This work has partially been funded by the Cluster of Excellence \textit{Multimodal Computing and Interaction} within the Excellence Initiative of the German Federal Government.

\end{document}